\documentclass{amsart}

\usepackage{amsmath}
\usepackage{amssymb}
\usepackage{amsthm}
\usepackage{tikz}
\usepackage{url}
\usepackage{ifthen}
\usepackage{enumitem}
\usepackage{subfig}
\usepackage{todonotes}
\usetikzlibrary{automata}
\usetikzlibrary{positioning}
\usetikzlibrary{calc}

\newcommand{\VAl}[2][]{\mathsf{VA}(\ifthenelse{\equal{#1}{}}{#2}{#2, #1})}

\newcommand{\Parikh}[1]{\Psi(#1)}

\newcommand{\monoidsubsetfont}[1]{\mathsf{#1}}
\newcommand{\Rio}{\monoidsubsetfont{R}}
\newcommand{\Lio}{\monoidsubsetfont{L}}

\newcommand{\Jo}{\monoidsubsetfont{J}}

\newcommand{\Ri}[1]{\Rio(#1)}
\newcommand{\Li}[1]{\Lio(#1)}

\newcommand{\J}[1]{\Jo(#1)}

\newcommand{\REG}{\mathsf{REG}}
\newcommand{\CF}{\mathsf{CF}}

\newcommand{\LFamily}{\mathcal{F}}

\DeclareMathOperator{\bin}{bin}

\newcommand{\F}{\mathcal{F}}
\newcommand{\C}{\mathcal{C}}
\newcommand{\Z}{\mathbb{Z}}
\newcommand{\B}{\mathbb{B}}
\newcommand{\N}{\mathbb{N}}
\newcommand{\M}[1][]{\mathbb{M}\ifthenelse{\equal{#1}{}}{}{(#1)}}

\newcommand{\Cfour}{\ensuremath{C_4}}
\newcommand{\Pfour}{\ensuremath{P_4}}

\newcommand{\defeq}{=}

\newcommand{\step}[1]{\Rightarrow_{#1}}

\newcommand{\congruence}{\equiv}

\newcommand{\emptyWord}{\lambda}

\newcommand{\labelfont}[1]{\mathsf{#1}}
\newcommand{\xn}{\labelfont{x_0}}
\newcommand{\xe}{\labelfont{x_1}}
\newcommand{\yn}{\labelfont{y_0}}
\newcommand{\ye}{\labelfont{y_1}}
\newcommand{\out}{\labelfont{x}}
\newcommand{\amalgtd}[1]{
\newcommand{\length}{2}
\begin{tikzpicture}[every state/.style={minimum size=10pt}]
\node[state, initial, initial text=, accepting]  (a) at (0,0) {?};
\path (a.center) ++(30:\length) node (0) [draw, state, accepting] {$0$};
\path (a.center) ++(-30:\length) node (1) [draw, state, accepting] {$1$};
\path[->] (a) edge node [above] {$\xn|\out$} (0)
          (a) edge node [below] {$\xe|\out$} (1)
          (0) edge [out=-100, in=100] node [left] {$\xe|\out$} (1)
          (1) edge [out=80, in=-80] node [right] {$\xn|\out$} (0)
          (a) edge [loop above] node {$\genfrac{}{}{0pt}{1}{\yn|\emptyWord}{\ye|\emptyWord}$} (a)
          (0) edge [loop right] node {$\genfrac{}{}{0pt}{1}{\yn|\emptyWord}{\ye|\emptyWord}$} (0)
          (1) edge [loop right] node {$\genfrac{}{}{0pt}{1}{\yn|\emptyWord}{\ye|\emptyWord}$} (1);
\end{tikzpicture}
}

\newcommand{\bztriangle}[1]{
\begin{tikzpicture}[every circle/.style={}, scale=#1]
\fill (0,0) circle (2pt) node (a) {}    (1,0) circle (2pt) node (b)  {}   (2,0) circle (2pt) node (c) {};
\draw (a.center) -- (b.center) -- (c.center);
\draw (a.center) ++(90:3pt) circle (3pt);
\draw (c.center) ++(90:3pt) circle (3pt);
\end{tikzpicture}
}

\newcommand{\bpair}[1]{
\begin{tikzpicture}[every circle/.style={}, scale=#1]
\fill (0,0) circle (2pt) node (a) {}    (1,0) circle (2pt) node (b)  {};
\draw (a.center) -- (b.center);
\end{tikzpicture}
}

\newcommand{\unloopedcycle}[1]{
\begin{tikzpicture}[every circle/.style={}, scale=#1]
\fill (0,0) circle (2pt) node (a) {}    (1,0) circle (2pt) node (b)  {}   (0,-1) circle (2pt) node (c) {}   (1,-1) circle (2pt) node (d) {};
\draw (a.center) -- (b.center) -- (d.center) -- (c.center) -- (a.center);
\end{tikzpicture}
}

\newcommand{\unloopedpath}[1]{
\begin{tikzpicture}[every circle/.style={}, scale=#1]
\fill (0,0) circle (2pt) node (a) {}    (1,0) circle (2pt) node (b)  {}   (2,0) circle (2pt) node (c) {}   (3,0) circle (2pt) node (d) {};
\draw (a.center) -- (b.center) -- (c.center) -- (d.center);
\end{tikzpicture}
}

\newlist{thmlist}{enumerate}{1}
\setlist[thmlist]{label=(\arabic*)}

\newenvironment{scorollary}{\begin{corollary}}{\end{corollary}}

\newtheorem{theorem}{Theorem}
\newtheorem{lemma}[theorem]{Lemma}
\newtheorem{corollary}[theorem]{Corollary}

\begin{document}
\title[Semlinearity and Context-Freeness]{Semilinearity and Context-Freeness of Languages Accepted by Valence Automata}
\author{P. Buckheister \and Georg Zetzsche}
\address{Fachbereich Informatik, Technische Universit\"{a}t Kaiserslautern, Postfach 3049, 67653 Kaiserslautern, Germany}

\begin{abstract}
Valence automata are a generalization of various models of automata with
storage.  Here, each edge carries, in addition to an input word, an element of
a monoid. A computation is considered valid if multiplying the monoid elements
on the visited edges yields the identity element. By choosing suitable
monoids, a variety of automata models can be obtained as special valence
automata. 

This work is concerned with the accepting power of valence automata.
Specifically, we ask for which monoids valence automata can accept only
context-free languages or only languages with semilinear Parikh image,
respectively. 

First, we present a characterization of those graph products (of monoids) for
which valence automata accept only context-free languages. Second, we provide a
necessary and sufficient condition for a graph product of copies of the
bicyclic monoid and the integers to yield only languages with semilinear Parikh
image when used as a storage mechanism in valence automata. Third, we show that
all languages accepted by valence automata over torsion groups have a
semilinear Parikh image.
\end{abstract}

\maketitle

\section{Introduction}
A valence automaton is a finite automaton in which each edge carries, in
addition to an input word, an element of a monoid. A computation is considered
valid if multiplying the monoid elements on the visited edges yields the
identity element. By choosing suitable monoids, one can obtain a wide range
of automata with storage mechanisms as special valence automata.  Thus, they
offer a framework for generalizing insights about automata with storage. 
For examples of automata as valence automata, see \cite{Gilman1996,Zetzsche2013b}.

In this work, we are concerned with the accepting power of valence automata.
That is, we are interested in relationships between the structure of the monoid
representing the storage mechanism and the class of languages accepted by the
corresponding valence automata. On the one hand, we address the question for
which monoids valence automata accept only \emph{context-free} languages. Since
the context-free languages constitute a very well-understood class, insights in
this direction promise to shed light on the acceptability of languages by
transferring results about context-free languages.

A very well-known result on context-free languages is Parikh's
Theorem~\cite{Parikh1966}, which states that the Parikh image (that is, the image
under the canonical morphism onto the free commutative monoid) of each
context-free language is semilinear (in this case, the language itself is also
called semilinear). It has various applications in proving that certain
languages are not context-free and its effective nature (one can actually
compute the semilinear representation) allows it to be used in decision
procedures for numerous problems (see \cite{LohreySteinberg2008} for an example
from group theory and \cite{KopczynskiTo2010} for others). It is therefore our
second goal to gain understanding about which monoids cause the corresponding
valence automata to accept only languages with a \emph{semilinear Parikh
image}. 

Our contribution is threefold. First, we obtain a characterization of those
graph products (of monoids) whose corresponding valence automata accept only
context-free languages.  Graph products are a generalization of the free and
the direct product in the sense that for each pair of participating factors, it
can be specified whether they should commute in the product.  Since valence
automata over a group accept only context-free languages if and only if the
group's word problem (and hence the group itself) can be described by a
context-free grammar, such a characterization had already been available for
groups in a result by Lohrey and S\'enizergues \cite{LohreySenizergues2007}.
Therefore, our characterization is in some sense an extension of Lohrey and
S\'enizergues' to monoids. 

Second, we present a necessary and sufficient condition for a graph product of
copies of the bicyclic monoid and the integers to yield, when used in valence
automata, only languages with semilinear Parikh image. Although this is a
smaller class of monoids than arbitrary graph products, it still covers a
number of storage mechanisms found in the literature, such as pushdown
automata, blind multicounter automata, and partially blind multicounter
automata (see \cite{Zetzsche2013b} for more information). Hence, our result is
a generalization of various semilinearity results about these types of
automata.

Third, we show that every language accepted by a valence automaton over a
torsion group has a semilinear Parikh image. On the one hand, this is
particularly interesting because of a result by Render \cite{Render2010}, which
states that for every monoid $M$, the languages accepted by valence automata
over $M$ either (1) coincide with the regular languages, (2) contain the blind
one-counter languages, (3) contain the partially blind one-counter languages,
or (4) are those accepted by valence automata over an infinite torsion group
(which is not locally finite). Hence, our result establishes a strong language
theoretic property in the fourth case and thus contributes to completing the
picture of language classes that can arise from valence automata.

On the other hand, Lohrey and Steinberg \cite{LohreySteinberg2008} have used
the fact that for certain groups, valence automata accept only semilinear
languages (in different terms, however) to obtain decidability of the rational
subset membership problem.  However, their procedures require that the
semilinear representation can be obtained effectively. Since there are torsion
groups where even the word problem is undecidable~\cite{Adian2010}, our result
yields examples of groups that have the semilinearity property but which do not
permit the computation of a corresponding representation. Our proof is based on
well-quasi-orderings (see, e.g., \cite{Kruskal1972}).

\section{Basic notions}
\label{sec:notions}
\newcommand{\emphasize}[1]{\emph{#1}}
We assume that the reader has some basic knowledge on formal languages and monoids.
In this section, we will fix some notation and introduce basic concepts.

A \emph{monoid} is a set $M$ together with an associative operation and a
neutral element.  Unless defined otherwise, we will denote the neutral element
of a monoid by $1$ and its operation by juxtaposition.  That is, for a monoid
$M$ and elements $a,b\in M$, $ab\in M$ is their product.
In each monoid $M$, we have the submonoids
\begin{eqnarray*}
\Ri{M}&\defeq& \{a\in M \mid \exists b\in M: ab=1 \}, \\
\Li{M}&\defeq& \{a\in M \mid \exists b\in M: ba=1 \}.
\end{eqnarray*}
When using a monoid $M$ as part of a control mechanism, the subset
 \[ \J{M}\defeq \{a\in M \mid \exists b,c\in M: bac=1\} \]
plays an important role\footnote{It should be noted that $\Ri{M}$, $\Li{M}$,
and $\J{M}$ are the $\mathcal{R}$-, $\mathcal{L}$-, and $\mathcal{J}$-class,
respectively, of the identity and hence are important concepts in the theory of
semigroups \cite{Howie1995}.} A \emph{subgroup} of a monoid is a subset
that is closed under the operation and is a group.

Let $\Sigma$ be a fixed countable set of abstract symbols, the finite subsets
of which are called \emph{alphabets}.  For a set of symbols $X\subseteq\Sigma$,
we will write $X^*$ for the set of words over $X$.  The empty word is denoted
by $\emptyWord\in X^*$.  Together with concatenation as its operation, $X^*$ is
a monoid. Given an alphabet $X$ and a monoid $M$, subsets of $X^*$ and
$X^*\times M$ are called \emph{languages} and \emph{transductions},
respectively. A \emph{family} is a set of languages that is closed under
isomorphism and contains at least one non-trivial member.
For a transduction $T\subseteq X^*\times Y^*$ and a language $L\subseteq X^*$,
we write $T(L)=\{v\in Y^* \mid \exists u\in L: (u,v)\in T\}$.
For any finite subset $S\subseteq M$ of a monoid, let $X_S$ be an alphabet
in bijection with $S$. Let $\varphi_S: X_S^*\to M$ be the morphism extending
this bijection. Then the set $\{w\in X_S^* \mid \varphi_S(w)=1 \}$ is called the
\emph{identity language of $M$ with respect to $S$}.

Let $\LFamily$ be a family of languages. An \emph{$\LFamily$-grammar} is a
quadruple $G=(N,T,P,S)$ where $N$ and $T$ are disjoint alphabets and $S\in
N$.  $P$ is a finite set of pairs $(A,M)$ with $A\in N$
and $M\subseteq (N\cup T)^*$, $M\in\LFamily$.  In this context, a pair $(A,M)\in P$
will also be denoted by $A\to M$.  We write $x\step{G} y$ if $x=uAv$ and
$y=uwv$ for some $u,v,w\in (N\cup T)^*$ and $(A,M)\in P$ with $w\in M$. The
\emph{language generated by $G$} is $L(G)=\{w\in T^* \mid S\step{G}^* w\}$. A
language $L$ is called \emph{algebraic over $\LFamily$} if there is an
$\LFamily$-grammar $G$ such that $L=L(G)$. The family of all languages that are
algebraic over $\LFamily$ is called the \emph{algebraic extension} of
$\LFamily$. The algebraic extension of the family of finite languages
is denoted $\CF$, its members are called \emph{context-free}.

Given an alphabet $X$, we write $X^\oplus$ for the set of maps $\alpha:X\to\N$.
Elements of $X^\oplus$ are called \emph{multisets}.  By way of pointwise
addition, written $\alpha+\beta$, $X^\oplus$ is a monoid. The \emph{Parikh
mapping} is the mapping $\Psi:X^*\to X^\oplus$ such that $\Psi(w)(x)$
is the number of occurrences of $x$ in $w$ for every $w\in X^*$ and
$x\in X$.

Let $A$ be a (not necessarily finite) set of symbols and $R\subseteq A^*\times
A^*$.  The pair $(A,R)$ is called a \emph{(monoid) presentation}. The smallest
congruence of $A^*$ containing $R$ is denoted by $\congruence_R$ and we will
write $[w]_R$ for the congruence class of $w\in A^*$.  The \emph{monoid
presented by $(A,R)$} is defined as $A^*/\mathord{\congruence_R}$. Note that
since we did not impose a finiteness restriction on $A$, every monoid has a
presentation. 

Let $M$ be a monoid. An \emph{automaton over $M$} is a tuple $A=(Q,M,E,q_0,F)$,
in which $Q$ is a finite set of \emph{states}, $E$ is a finite subset of
$Q\times M\times Q$ called the set of \emph{edges}, $q_0\in Q$ is the
\emph{initial state}, and $F\subseteq Q$ is the set of \emph{final states}. The
\emph{step relation} $\step{A}$ of $A$ is a binary relation on $Q\times M$, for
which $(p,a) \step{A} (q,b)$ if and only if there is an edge $(p,c,q)$ such that $b=ac$.
The set \emph{generated by $A$} is then $S(A)\defeq\{a\in M \mid \exists q\in F:
(q_0,1)\step{A}^* (q,a) \}$. A set $R\subseteq M$ is called \emph{rational} if
it can be written as $R=S(A)$ for some automaton $A$ over $M$.  Rational
languages are also called \emph{regular}, the corresponding class is denoted
$\REG$.  A class $\C$ for which $L\in\C$ implies $T(L)\in\C$ for every
rational transduction $T$ is called a \emph{full trio}.

For $n\in\N$ and $\alpha\in X^\oplus$, we use $n\alpha$ to denote
$\alpha+\cdots+\alpha$ ($n$ summands).  A subset $S\subseteq X^\oplus$ is
\emph{linear} if there are elements $\alpha_0,\ldots,\alpha_n$ such that
$S=\{\alpha_0+\sum_{i=1}^n m_i\alpha_i \mid m_i\in\N,~ 1\le i\le n\}$.  A set
$S\subseteq C$ is called \emph{semilinear} if it is a finite union of linear
sets.  In slight abuse of terminology, we will sometimes call a language $L$
semilinear if the set $\Parikh{L}$ is semilinear. 

A \emph{valence automaton over $M$} is an automaton $A$ over $X^*\times M$,
where $X$ is an alphabet. Instead of $A=(Q,X^*\times M, E,q_0,F)$, we then also
write $A=(Q,X,M,E,q_0,F)$ and for an edge $(p,(w,m),q)\in E$, we also write
$(p,w,m,q)$. The \emph{language accepted by $A$} is defined as $L(A)\defeq
\{w\in X^* \mid (w,1)\in S(A)\}$.  The class of languages accepted by valence
automata over $M$ is denoted by $\VAl{M}$. It is well-known that $\VAl{M}$ is
the smallest full trio containing every identity language of $M$ (see, for
example, \cite{Kambites2009}).

A \emph{graph} is a pair $\Gamma=(V,E)$ where $V$ is a finite set and
$E\subseteq \{S\subseteq V \mid 1\le|S|\le 2\}$.  The elements of $V$ are
called \emph{vertices} and those of $E$ are called \emph{edges}.  If $\{v\}\in
E$ for some $v\in V$, then $v$ is called a \emph{looped} vertex, otherwise it
is \emph{unlooped}.  A \emph{subgraph} of $\Gamma$ is a graph $(V',E')$ with
$V'\subseteq V$ and  $E'\subseteq E$.  Such a subgraph is called \emph{induced
(by $V'$)} if $E'=\{ S\in E \mid S\subseteq V'\}$, i.e. $E'$ contains all edges
from $E$ incident to vertices in $V'$.  By $\Gamma\setminus\{v\}$, for $v\in
V$, we denote the subgraph of $\Gamma$ induced by $V\setminus \{v\}$.  Given a
graph $\Gamma=(V,E)$, its \emph{underlying loop-free graph} is $\Gamma'=(V,E')$
with $E'=E\cap \{S\subseteq V \mid |S|=2 \}$.  For a vertex $v\in V$, the
elements of $N(v)=\{w\in V \mid \{v,w\}\in E\}$ are called \emph{neighbors} of
$v$. Moreover, a \emph{clique} is a graph in which any two distinct vertices
are adjacent. A \emph{simple path of length $n$} is a sequence $x_1,\ldots,x_n$
of pairwise distinct vertices such that $\{x_i,x_{i+1}\}\in E$ for $1\le i<n$.
If, in addition, we have $\{x_n, x_1\}\in E$, it is called a \emph{cycle}.
Such a cycle is called \emph{induced} if $\{x_i, x_j\}\in E$ implies $|i-j|=1$
or $\{i,j\}=\{1,n\}$. A loop-free graph $\Gamma=(V,E)$ is \emph{chordal}
if it does not contain an induced cycle of length $\ge 4$.  It is well-known
that every chordal graph contains a vertex whose neighborhood is a clique
\cite{Dirac1961}. By \Cfour{} and \Pfour{}, we denote the cycle of length 4 and
the simple path of length 4, respectively (see figures \ref{fig:cfour} and
\ref{fig:pfour}).  A loop-free graph is called a \emph{transitive forest} if it
is the disjoint union of comparability graphs of rooted trees. A result by
Wolk~\cite{Wolk1965} states that a graph is a transitive forest if and only if
it contains neither \Cfour{} nor \Pfour{} as an induced subgraph.

\begin{figure}[t]
\centering
\subfloat[]{\label{fig:cfour}\unloopedcycle{1}}
\subfloat[]{\label{fig:pfour}\unloopedpath{1}}
\caption{Graphs \Cfour{} and \Pfour{}.}
\end{figure}

\newcommand{\restrict}[2]{#1\mathord{\upharpoonright_{#2}}}
Let $\Gamma=(V,E)$ be a loop-free graph and $M_v$ a monoid for each $v\in V$
with a presentation $(A_v, R_v)$ such that the $A_v$ are pairwise disjoint.
Then the \emph{graph product} $M=\M[\Gamma, (M_v)_{v\in V}]$ is the monoid given
by the presentation $(A, R)$, where $A=\bigcup_{v\in V} A_v$ and \[ R=\{(ab,ba)
\mid a\in A_v, b\in A_w, \{v,w\}\in E\}\cup\bigcup_{v\in V} R_v. \] Note that
for each $v\in V$, there is a map $\varphi_v:M\to M_v$ such that $\varphi_v$ is
the identity map on $M_v$. When $V=\{0,1\}$ and $E=\emptyset$, we also write
$M_0 * M_1$ for $M$ and call this the \emph{free product} of $M_0$ and $M_1$.
Given a subset $U\subseteq V$, we write $\restrict{M}{U}$ for the product
$\M[\Gamma', (M_v)_{v\in U}]$, where $\Gamma'$ is the subgraph induced by $U$.

By $\B$, we denote the monoid presented by $(A,R)$ with $A=\{x,\bar{x}\}$ and
$R=(x\bar{x}, \emptyWord)$. The elements $[x]_R$ and $[\bar{x}]_R$ are called
its \emph{positive} and \emph{negative generator}, respectively. The set $D_1$
of all $w\in\{x,\bar{x}\}^*$ with $[w]_R=[\emptyWord]_R$ is called the
\emph{Dyck language}. The group of integers is denoted $\Z$.  Here, we call
$1\in\Z$ its \emph{positive} and $-1\in\Z$ its \emph{negative generator}.

Let $\Gamma=(V,E)$ be a (not necessarily loop-free) graph. Furthermore, for each
$v\in V$, let $M_v$ be a copy of $\B$ if $v$ is an unlooped vertex and a copy
of $\Z$ if $v$ is looped. If $\Gamma^-$ is obtained from $\Gamma$ by removing
all loops, we write $\M\Gamma$ for the graph product $\M[\Gamma^-, (M_v)_{v\in
V}]$.  For information on valence automata over monoids $\M\Gamma$, see
\cite{Zetzsche2013b}.

For $i\in\{0,1\}$, let $M_i$ be a monoid and let $\varphi_i:N\to M_i$ be an
injective morphism.  Let $\equiv$ be the smallest congruence in $M_0*M_1$ such
that $\varphi_0(a)\equiv\varphi_1(a)$ for every $a\in N$.  Then the monoid
$(M_0*M_1)/\equiv$ is denoted by $M_0 *_N M_1$ and called a \emph{free product
with amalgamation}.

\section{Auxiliary Results}\label{sec:aux}

In this section, we present auxiliary results that are used in later sections.
In the following, we will call a monoid $M$ an \emph{FRI-monoid} (or say that
$M$ has the FRI-property) if for every finitely generated submonoid $N$ of $M$,
the set $\Ri{N}$ is finite.  In \cite{Render2010} and independently in
\cite{Zetzsche2011b}, the following was shown.
\begin{theorem}\label{thm:regfri}
For each monoid $M$, the following are equivalent:
\begin{thmlist}
\item $M$ is an FRI-monoid.
\item $\VAl{M}=\REG$.
\end{thmlist}
\end{theorem}

The first two lemmas state well-known facts from semigroup theory for which we
provide short proofs for the sake of accessibility.
\begin{lemma}\label{lemma:grouporbicyclic}
For each monoid $M$, exactly one of the following holds:
\begin{thmlist}
\item $\J{M}$ is a group,
\item $M$ contains a copy of $\B$ as a submonoid.
\end{thmlist}
\end{lemma}
\begin{proof}
If $\Ri{M}=\Li{M}$, then $\J{M}=\Ri{M}=\Li{M}$ and hence $\J{M}$ is a group.
Otherwise, if $x\in\Ri{M}\setminus\Li{M}$ with $xy=1$, it can be verified
straightforwardly that the submonoid generated by $x$ and $y$ is isomorphic to
$\B$. If $\Li{M}\setminus\Ri{M}\ne\emptyset$, we can proceed analogously.  The
two cases are mutually exclusive, since in the second case, we have $xy=1$ and
$yx\ne 1$, where $x$ and $y$ are the positive and negative generator of $\B$,
respectively.  This, however, cannot happen in a group.
\end{proof}

\begin{lemma}\label{lemma:finiteorinfinite}
For each monoid $M$, exactly one of the following holds:
\begin{thmlist}
\item $M$ is an FRI-monoid.
\item There is a finitely generated submonoid $N\subseteq M$ and infinite
subsets $S\subseteq\Ri{N}$, $S'\subseteq\Li{N}$ such that (i) no two distinct
elements of $S$ have a right inverse in common and (ii) no two distinct elements
of $S'$ have a left inverse in common.
\end{thmlist}
\end{lemma}
\begin{proof}
The conditions are clearly mutually exclusive.  If $M$ is not an FRI-monoid, it
has a finitely generated submonoid $N$ with infinite $\Ri{N}$. 
Distinguishing the cases of Lemma \ref{lemma:grouporbicyclic} for $N$ yields
the required sets. 
\end{proof}

We will employ a result by van Leeuwen~\cite{vanLeeuwen1974} that generalizes
Parikh's theorem.  It states that semilinearity of all languages is preserved
by building the algebraic extension of a language family.

\begin{theorem}[van Leeuwen]\label{thm:algsemilinear}
Let $\LFamily$ be a family of semilinear languages. Then every language that is
algebraic over $\LFamily$ is also semilinear.
\end{theorem}

In light of the previous theorem, the following implies that the class of
monoids $M$ for which $\VAl{M}$ contains only semilinear languages is closed
under taking free products with amalgamation over a finite identified subgroup
that contains the identity of each factor.
In the case where the factors are residually finite groups, this was already
shown in \cite[Lemma 8]{LohreySteinberg2008} (however, for a more general
operation than free products with amalgamation). The following also implies
that if $\VAl{M_i}$ contains only context-free languages for $i\in\{0,1\}$,
then this is also true for $\VAl{M_0*_FM_1}$. This is due to the fact that
clearly, the class of context-free languages is its own algebraic extension.
\begin{theorem}\label{thm:amalgalgebraic}
For each $i\in\{0,1\}$, let $M_i$ be a finitely generated monoid and $F$ be a
subgroup that contains $M_i$'s identity.  Then every language in $\VAl{M_0 *_F
M_1}$ is algebraic over $\VAl{M_0}\cup\VAl{M_1}$.
\end{theorem}
\begin{proof}
Since the algebraic extension of a full trio is again a full trio, it suffices
to show that with respect to some generating set $S\subseteq M_0 *_F M_1$, the
identity language of $M_0 *_F M_1$ is algebraic over $\VAl{M_0}\cup\VAl{M_1}$.

For $i\in\{0,1\}$, let $S_i\subseteq M_i$ be a finite generating set for $M_i$
such that $F\subseteq S_i$.  Furthermore, let $X_i$ be an alphabet in bijection
with $S_i$ and let $\varphi_i:X_i^*\to M_i$ be the morphism extending this
bijection. Moreover, let $Y_i\subseteq X_i$ be the subset with
$\varphi_i(Y_i)=F$. Let $\psi_i: M_i\to M_0 *_F M_1$ be the canonical morphism.
Since $F$ is a subgroup of $M_0$ and $M_1$, $\psi_0$ and $\psi_1$ are injective
(see e.g. \cite[Theorem 8.6.1]{Howie1995}). Let $X=X_0\cup X_1$ and let
$\varphi:X^*\to M_0 *_F M_1$ be the morphism extending $\psi_0\varphi_0$ and
$\psi_1\varphi_1$. Then the identity language of $M_0*_F M_1$ is $\varphi^{-1}(1)$
and we shall prove the theorem by showing that $\varphi^{-1}(1)$ is algebraic
over $\VAl{M_0}\cup\VAl{M_1}$. 
We will make use of the following fact about free products with amalgamation of
monoids with a finite identified subgroup.  Let
$s_1,\ldots,s_n,s'_1,\ldots,s'_m\in (X_0^*\setminus\varphi_0^{-1}(F))\cup
(X_1^*\setminus \varphi_1^{-1}(F))$, such that $s_j\in X_i^*$ if and only if
$s_{j+1}\in X_{1-i}^*$ for $1\le j<n$, $i\in\{0,1 \}$ and $s'_j\in X_i^*$ if
and only if $s'_{j+1}\in X_{1-i}^*$ for $1\le j<m$, $i\in\{0,1\}$. Then the
equality $\varphi(s_1\cdots s_n)=\varphi(s'_1\cdots s'_m)$ implies $n=m$.  A
stronger statement was shown in \cite[Lemma 10]{LohreySenizergues2008}.
We will refer to this as the \emph{syllable property}.

For each $i\in\{0,1\}$ and $f\in F$, we define $L_{i,f}=\varphi_i^{-1}(f)$ and
write $y_f$ for the symbol in  $Y_i$ with $\varphi_i(y_f)=f^{-1}$. Then clearly
$L_{i,1}\in\VAl{M_i}$. Furthermore, since 
\[ L_{i,f}=\{w\in X_i^* \mid y_fw\in L_{i,1} \}, \]
(here we again use that $F$ is a group) we can obtain $L_{i,f}$ from $L_{i,1}$
by a rational transduction and hence $L_{i,f}\in \VAl{M_i}$.

Let $\F=\VAl{M_0}\cup\VAl{M_1}$.  Since for each $\F$-grammar $G$, it is
clearly possible to construct an $\F$-grammar $G'$ such that $L(G')$ consists
of all sentential forms of $G$, it suffices to construct an $\F$-grammar
$G=(N,T,P,S)$ with $N\cup T=X$ and $S\step{G}^* w$ if and only if
$\varphi(w)=1$ for $w\in X^*$. We construct $G=(N,T,P,S)$ as follows. Let
$N=Y_0\cup Y_1$ and $T=(X_0\cup X_1)\setminus (Y_0\cup Y_1)$.  As productions,
we have $y\to L_{1-i, f}$ for each $y\in Y_i$ where $f=\varphi_i(y)$. Since
$1\in F$, we have an $e_i\in Y_i$ with $\varphi_i(e_i)=1$.  As the start
symbol, we choose $S=e_0$.  We claim that for $w\in X^*$, we have $S\step{G}^*
w$ if and only if $\varphi(w)=1$.  

The ``only if'' is clear. Thus, let $w\in X^*$ with $\varphi(w)=1$. We write
$w=w_1\cdots w_n$ such that $w_j\in X_0^*\cup X_1^*$ for all $1\le j\le n$ such
that $w_j\in X_i^*$ if and only if $w_{j+1}\in X_{1-i}^*$ for $i\in\{0,1\}$ and
$1\le j<n$.  We show by induction on $n$ that $S\step{G}^* w$. For $n\le 1$, we
have $w\in X_i^*$ for some $i\in\{0,1\}$.  Since
$1=\varphi(w)=\psi_i(\varphi_i(w))$ and $\psi_i$ is injective, we have
$\varphi_i(w)=1$ and hence $w\in L_{i,1}$.  This means $S=e_0\step{G}w$ or
$S=e_0\step{G}e_1\step{G}w$, depending on whether $i=1$ or $i=0$.

Now let $n\ge 2$. We claim that there is a $1\le j\le n$ with $\varphi(w_j)\in
F$. Indeed, if $\varphi(w_j)\notin F$ for all $1\le j\le n$ and since
$\varphi(w_1\cdots w_n)=1=\varphi(\emptyWord)$, the syllable property implies
$n=0$, against our assumption. Hence, let $f=\varphi(w_j)\in F$.  Furthermore,
let $w_j\in X_i^*$ and choose $y\in Y_{1-i}$ so that $\varphi_{1-i}(y)=f$. Then
$\psi_i(\varphi_i(w_j))=\varphi(w_j)=f$ and the injectivity of $\psi_i$ yields
$\varphi_i(w_j)=f$. Hence, $w_j\in L_{i,f}$ and thus $w'=w_1\cdots
w_{j-1}yw_{j+1}\cdots w_n \step{G} w$. For $w'$ the induction hypothesis holds,
meaning $S\step{G}^* w'$ and thus $S\step{G}^* w$.
\end{proof}

\section{Context-Freeness}\label{sec:cf}
In this section, we are concerned with the context-freeness of languages
accepted by valence automata over graph products. The first lemma is a simple
observation and we will not provide a proof. In the case of groups, it appeared
in \cite{Green1990}.
\begin{lemma}\label{lemma:neighbors}
Let $\Gamma=(V,E)$ and $M=\M[\Gamma, (M_v)_{v\in V}]$ be a graph product. Then
for each $v\in V$
\[ M \cong (\restrict{M}{V\setminus \{v\}})*_{\restrict{M}{N(v)}} (\restrict{M}{N(v)}\times M_v). \]
\end{lemma}

The following is a result by Lohrey and S\'{e}nizergues
\cite{LohreySenizergues2007}.  A finitely generated group is called
\emph{virtually free} if it has a free subgroup of finite index.
\begin{theorem}[Lohrey, S\'{e}nizergues]\label{thm:gpgroup}
Let $G_v$ be a finitely generated non-trivial group for each $v\in V$. Then $\M[\Gamma, (G_v)_{v\in V}]$ is
virtually free if and only if
\begin{thmlist}[after*=\label{gcfcond:last}]
\item\label{gcfcond:cf} for each $v\in V$, $G_v$ is virtually free,
\item\label{gcfcond:direct} if $G_v$ and $G_w$ are infinite and $v\ne w$, then
$\{v,w\}\notin E$,
\item\label{gcfcond:triangle} if $G_v$ is infinite, $G_u$ and $G_w$ are finite and
$\{v,u\},\{v,w\}\in E$, then $\{u,w\}\in E$, and
\item\label{gcfcond:chordal} the graph $\Gamma$ is chordal.
\end{thmlist}
\end{theorem}

In order to prove that certain languages are not context-free, we will employ
the following well-known Iteration Lemma by Ogden \cite{Ogden1968}.
\begin{lemma}[Ogden]\label{lemma:ogden}
For each context-free language $L$, there is an integer $m$ such that for
any word $z\in L$ and any choice of at least $m$ distinct marked positions
in $z$, there is a decomposition $z=uvwxy$ such that:
\begin{thmlist}
\item\label{ogden:wgeone} $w$ contains at least one marked position.
\item\label{ogden:either} Either $u$ and $v$ both contain marked positions, or
$x$ and $y$ both contain marked positions.
\item\label{ogden:most} $vwx$ contains at most $m$ marked positions.
\item\label{ogden:pump} $uv^iwx^iy\in L$ for every $i\ge 0$.
\end{thmlist}
\end{lemma}

Aside from Theorem \ref{thm:amalgalgebraic}, the following is the key tool to
prove our result on context-freeness. We call a monoid $M$ \emph{context-free} if
$\VAl{M}\subseteq\CF$.
\begin{lemma}\label{lemma:directproduct}
The direct product of monoids $M_0$ and $M_1$ is context-free if and only if
for some $i\in\{0,1\}$, $M_i$ is context-free and $M_{1-i}$ is an FRI-monoid.
\end{lemma}
\begin{proof}
Suppose $M_i$ is context-free and $M_{1-i}$ is an FRI-monoid. Then each language
$L\in\VAl{M_i\times M_{1-i}}$ is contained in $\VAl{M_i\times N}$ for some
finitely generated submonoid $N$ of $M_{1-i}$.  Since $M_{1-i}$ is
an FRI-monoid, $N$ has finitely many right-invertible elements and hence $\J{N}$
is a finite group.  Since no element outside of $\J{N}$ can appear in a product
yielding the identity, we may assume that $L\in\VAl{M_i\times\J{N}}$.  This
means, however, that $L$ can be accepted by a valence automaton over $M_i$ by
keeping the right component of the storage monoid in the state of the
automaton. Hence, $L\in\VAl{M_i}$ is context-free.

Suppose $\VAl{M_0\times M_1}\subseteq\CF$. Then certainly
$\VAl{M_i}\subseteq\CF$ for each $i\in\{0,1\}$.  This means we have to show
that at least one of the monoids $M_0$ and $M_1$ is an FRI-monoid and thus,
toward a contradiction, assume that none of them is. We provide two proofs for
the fact that $\VAl{M_0\times M_1}$ contains non-context-free languages in this
case. One is very short and the other is elementary in the sense that it does
not invoke the fact that context-free groups are virtually free.

\emph{First proof.} By Lemma \ref{lemma:grouporbicyclic}, for each $i$, either
$\J{M_i}$ is an infinite subgroup of $M_i$ or $M_i$ contains a copy of $\B$ as
a submonoid.  Since every infinite virtually free group contains an element of
infinite order, we have that for each $i$, either (1) $\J{M_i}$ is an infinite
group and hence contains a copy of $\Z$ or (2) $M_i$ contains a copy of $\B$.
In any case, $\VAl{M_0\times M_1}$ contains the language $\{a^nb^mc^nd^m \mid
n,m\ge 0\}$, which is not context-free.

\emph{Second proof.} By Lemma \ref{lemma:finiteorinfinite}, for each $i$, there
is a finitely generated submonoid $N_i\subseteq M_i$ and infinite sets 
$S_0\subseteq \Ri{N_0}$ and $S_1\subseteq\Li{N_1}$ such that the elements of $S_0$ have
pairwise disjoint sets of right inverses in $N_0$ and the elements of $S_1$
have pairwise disjoint sets of left inverses in $N_1$. Let $X_i$ be an alphabet
large enough that we can find a surjective morphism $\varphi_i:X_i^*\to N_i$
for each $i\in\{0,1\}$.  Furthermore, let $\#$ be a symbol with $\#\notin
X_0\cup X_1$. The language  
\[ L=\{r_0\#r_1\#s_0\#s_1 \mid r_i,s_i\in X_i^*,~\varphi_i(r_is_i)=1~\text{for each $i\in\{0,1\}$}\} \]
is clearly contained in $\VAl{M_0\times M_1}$. We shall use the Iteration Lemma
to show that $L$ is not context-free. Suppose $L$ is context-free and let $m$
be the constant provided by Lemma \ref{lemma:ogden}.  For each $a\in \Ri{N_0}$,
let $\ell_0(a)$ be the minimal length of a word $w\in X_0^*$ with
$a\varphi_0(w)=1$. Furthermore, for $a\in \Li{N_1}$, let $\ell_1(a)$ be the
minimal length of a word $w\in X_1^*$ with $\varphi_1(w)a=1$. The existence of
the sets $S_0$ and $S_1$ guarantees that there are $a_0\in \Ri{N_0}$ and
$a_1\in\Li{N_1}$ such that $\ell_0(a_0)\ge m$ and $\ell_1(a_1)\ge m$.  Choose
$r_0\in X_0^*$ and $s_1\in X_1^*$ such that $\varphi_0(r_0)=a_0$ and
$\varphi_1(s_1)=a_1$. Furthermore, let $r_1\in X_1^*$ be a word of minimal
length among those with $\varphi_1(r_1s_1)=1$ and let $s_0\in X_0^*$ be a word
of minimal length among those with $\varphi_0(r_0s_0)=1$. These choices
guarantee $|r_1|\ge m$ and $|s_0|\ge m$. Moreover, the word $z=r_0\#r_1\#s_0\#s_1$ is in $L$.

Let $z=uvwxy$ be the decomposition provided by the Iteration Lemma, where we
choose the positions in the subword $r_1\#s_0$ to be marked. In the following,
we call $r_0,r_1,s_0,s_1$ the \emph{segments} of the word $z$. Clearly, $v$ and
$x$ cannot contain the symbol $\#$. Therefore, by Condition \ref{ogden:either},
at least one of the words $v$ and $x$ lies in one of the middle segments. By
Condition \ref{ogden:most}, they have to lie in the same segment or in
neighboring segments. Hence, we have two cases:
\begin{itemize}
\item If $v$ or $x$ lies in the segment $r_1$, none of them lies in $s_1$.
Thus, by pumping with $i=0$, we obtain a word $r'_0\#r'_1\#s'_0\#s_1\in L$ with
$|r'_1|<|r_1|$ and $\varphi_1(r'_1s_1)=1$, contradicting the choice of $r_1$.
\item If $v$ or $x$ lies in the segment $s_0$, none of them lies in $r_0$.
Thus, by pumping with $i=0$, we obtain a word $r_0\#r'_1\#s'_0\#s'_1\in L$ with
$|s'_0|<|s_0|$ and $\varphi_1(r_0s'_0)=1$, contradicting the choice of $s_0$.
\end{itemize}
This proves that $L$ is not context-free and hence the lemma.
\end{proof}

We are now ready to prove our main result on context-freeness. Since for a
graph product $M=\M[\Gamma, (M_v)_{v\in V}]$, there is a morphism
$\varphi_v:M\to M_v$ for each $v\in V$ that restricts to the identity on $M_v$, we have
$\J{M}\cap M_v=\J{M_v}$: While the inclusion ``$\supseteq$'' is true for any
submonoid, given $b\in\J{M}\cap M_v$ with $abc=1$, $a,c\in M$, we also have
$\varphi_v(a)b\varphi_v(c)=\varphi_v(abc)=1$ and hence $b\in \J{M_v}$. This means
no element of $M_v\setminus\J{M_v}$ can appear in a product yielding the identity.
In particular, removing a vertex $v$ with $\J{M_v}=\{1\}$ will not change $\VAl{M}$.
Hence, our requirement that $\J{M_v}\ne\{1\}$ is not a serious restriction.
\begin{theorem}\label{thm:cf}
Let $\Gamma=(V,E)$ and let $\J{M_v}\ne\{1\}$ for any $v\in V$. $M=\M[\Gamma,
(M_v)_{v\in V}]$ is context-free if and only if
\begin{thmlist}[after*=\label{cfcond:last}]
\item\label{cfcond:cf} for each $v\in V$, $M_v$ is context-free,
\item\label{cfcond:direct} if $M_v$ and $M_w$ are not FRI-monoids and $v\ne
w$, then $\{v,w\}\notin E$,
\item\label{cfcond:triangle} if $M_v$ is not an FRI-monoid, $M_u$ and $M_w$
are FRI-monoids and $\{v,u\},\{v,w\}\in E$, then $\{u,w\}\in E$, and
\item\label{cfcond:chordal} the graph $\Gamma$ is chordal.
\end{thmlist}
\end{theorem}
\begin{proof}
First, we show that conditions (1)--\ref{cfcond:last} are necessary.  For
\ref{cfcond:cf}, this is immediate and for \ref{cfcond:direct}, this follows
from Lemma \ref{lemma:directproduct}.  If \ref{cfcond:triangle} is violated
then for some $u,v,w\in V$, $M_v\times (M_u*M_w)$ is a submonoid of $M$ such
that $M_u$ and $M_w$ are FRI-monoids and $M_v$ is not. Since $M_u$ and $M_w$
contain non-trivial (finite) subgroups, $M_u*M_w$ contains an infinite group
and is thus not an FRI-monoid, meaning $M_v\times (M_u*M_w)$ is not
context-free by Lemma \ref{lemma:directproduct}. 

Suppose \ref{cfcond:chordal} is violated for context-free $M$. By
\ref{cfcond:direct} and \ref{cfcond:triangle}, any induced cycle of length at
least four involves only vertices with FRI-monoids. Each of these,
however, contains a non-trivial finite subgroup. This means $M$ contains an
induced cycle graph product of non-trivial finite groups, which is not
virtually free by Theorem \ref{thm:gpgroup} and hence has a non-context-free
identity language.

In order to prove the other direction, we note that $\VAl{M}\subseteq\CF$
follows if $\VAl{M'}\subseteq\CF$ for every finitely generated submonoid
$M'\subseteq M$. Since every such submonoid is contained in a graph product
$N=\M[\Gamma, (N_v)_{v\in V}]$ where each $N_v$ is a finitely generated
submonoid of $M_v$, it suffices to show that for such graph products, we have
$\VAl{N}\subseteq\CF$. This means whenever $M_v$ is an FRI-monoid, $N_v$ has
finitely many right-invertible elements. Moreover, since
$N_v\cap\J{N}=\J{N_v}$, no element of $N_v\setminus\J{N_v}$ can appear in a
product yielding the identity.  Hence, if $N_v$ is generated by $S\subseteq
N_v$, replacing $N_v$ by the submonoid generated by $S\cap\J{N_v}$ does not
change the identity languages of the graph product. Thus, we assume that each
$N_v$ is generated by a finite subset of $\J{N_v}$. Therefore, whenever $M_v$
is an FRI-monoid, $N_v$ is a finite group.

We first establish sufficiency in the case that $M_v$ is an FRI-monoid for
every $v\in V$ and proceed by induction on $|V|$.  This means that $N_v$ is a
finite group for every $v\in V$.  Since $\Gamma$ is chordal, there is a $v\in
V$ whose neighborhood is a clique.  This means $\restrict{N}{N(v)}$ is a finite
group and hence $\restrict{N}{N(v)}\times N_v$ context-free by Lemma
\ref{lemma:directproduct}.  Since $\restrict{N}{V\setminus\{v\}}$ is
context-free by induction, Theorem \ref{thm:amalgalgebraic} and Lemma
\ref{lemma:neighbors} imply that $N$ is context-free.

To complete the proof, suppose there are $n$ vertices $v\in V$ for which $M_v$
is not an FRI-monoid. We proceed by induction on $n$. The case $n=0$ is
treated above. Choose $v\in V$ such that $M_v$ is not an FRI-monoid. For each
$u\in N(v)$, $M_u$ is an FRI-monoid by condition \ref{cfcond:direct}, and
hence $N_u$ a finite group.  Furthermore, condition \ref{cfcond:triangle}
guarantees that $N(v)$ is a clique and hence $\restrict{N}{N(v)}$ is a finite
group.  As above, Theorem \ref{thm:amalgalgebraic} and Lemma
\ref{lemma:neighbors} imply that $N$ is context-free.
\end{proof}

\begin{corollary}
Let $\Gamma=(V,E)$. Then $\VAl{\M[\Gamma, (M_v)_{v\in V}]}\subseteq\CF$ if and
only if
\begin{thmlist}
\item for each $v\in V$, $\VAl{M_v}\subseteq\CF$,
\item if $\REG\subsetneq\VAl{M_v}$ and $\REG\subsetneq\VAl{M_w}$ and $v\ne w$,
then $\{v,w\}\notin E$,
\item if $\REG\subsetneq\VAl{M_v}$, $\VAl{M_u}=\VAl{M_w}=\REG$ and $\{v,u\}\in
E$ and $\{v,w\}\in E$, then $\{u,w\}\in E$, and
\item the graph $\Gamma$ is chordal.
\end{thmlist}
\end{corollary}

\section{Semilinearity}\label{sec:semilinearity}
A well-known theorem by Chomsky and Sch\"{u}tzenberger
\cite{ChomskySchutzenberger1963} was re-proved and phrased in terms of valence
automata in the following way by Kambites \cite{Kambites2009}.
\begin{theorem}\label{thm:freecf}
$\VAl{\Z*\Z}=\CF$.
\end{theorem}

The next lemma can be shown using standard methods of formal language theory.
See \cite{LohreySteinberg2008,Zetzsche2013b} for a proof.
\begin{lemma}\label{lemma:prodz}
Let $M$ be a monoid such that all languages in $\VAl{M}$ are semilinear. Then
every languages in $\VAl{M\times\Z}$ is semilinear.
\end{lemma}

By a simple product construction, one can show the following.
\begin{lemma}\label{lemma:prodinclusion}
If $\VAl{N_i}\subseteq\VAl{M_i}$ for $i=0,1$,  then $\VAl{N_0\times
N_1}\subseteq\VAl{M_0\times M_1}$.
\end{lemma}

\begin{lemma}\label{lemma:bbnonsemilinear}
$\VAl{\B\times\B}$ contains a non-semilinear language.
\end{lemma}
\begin{proof}
$\VAl{\B\times\B}$ is the class of languages accepted by partially blind
two-counter machines \cite{Zetzsche2013b}.  Greibach \cite{Greibach1978} and,
independently, Jantzen \cite{Jantzen1979a} have shown that such machines can
accept the language $L_1=\{wc^n \mid w\in \{0,1\}^*,~ n\le \bin(w) \}$, where
$\bin(w)$ denotes the number obtained by interpreting $w$ as a base $2$
representation: $\bin(w1)=2\cdot \bin(w)+1$, $\bin(w0)=2\cdot \bin(w)$,
$\bin(\emptyWord)=0$.  This means $L_1\cap\{1\}\{0,c\}^*=\{10^nc^m \mid m\le
2^n\}$ is also in $\VAl{\B\times\B}$, which is clearly not semilinear.
\end{proof}

The next result also appears in \cite{Zetzsche2013b}, where, however, it
was not made explicit that the undecidable language is unary.
\begin{lemma}\label{lemma:pfour}
If $\Gamma$'s underlying loop-free graph contains $\Pfour{}$ as an induced
subgraph, then $\VAl{\M\Gamma}$ contains an undecidable unary language.
\end{lemma}
\begin{proof}
Let $\Gamma=(V,E)$ and $\mathring{\Gamma}$ be the graph obtained from $\Gamma$
by adding a loop to every unlooped vertex.  For notational reasons, we assume
that the vertex set of $\mathring{\Gamma}$ is $\mathring{V}=\{\mathring{v}\mid
v\in V\}$.  Recall that $\M\Gamma$ is defined as $\M(\Gamma^-, (M_v)_{v\in
V})$, where $M_v$ is $\Z$ or $\B$, depending on whether $v$ is looped or not.
In the following, we write $a_v$ and $\bar{a}_v$ for $M_v$'s positive and
negative generator, respectively.
Lohrey and Steinberg \cite{LohreySteinberg2008} show that there are
rational sets $\mathring{R},\mathring{S}\subseteq\M\mathring{\Gamma}$ over
positive generators such that for a certain $\mathring{u}\in \mathring{V}$, given $n\in\N$, it is
undecidable whether $1\in a_{\mathring{u}}^n\mathring{R}\mathring{S}^{-1}$. Note that the
morphism $\varphi:\M\Gamma\to\M\mathring{\Gamma}$ with
$\varphi(a_{v})=a_{\mathring{v}}$ and
$\varphi(\bar{a}_v)=\bar{a}_{\mathring{v}}$ induces an isomorphism between the submonoids
generated by positive generators and between the submonoids generated by the negative generators.
Thus, we find rational sets
$R,S\subseteq\M\Gamma$ over positive generators with $\varphi(R)=\mathring{R}$ and
$\varphi(S)=\mathring{S}$.

If $w$ is a word over positive generators in $\M\Gamma$, $w=a_1\cdots a_n$, then
we let $\bar{w}=\bar{a}_n\cdots\bar{a}_1$. This is well-defined, for if $a_1\cdots a_n=b_1\cdots b_m$,
for positive generators $a_1,\ldots,a_n,b_1,\ldots,b_m$ then $\varphi(a_1\cdots a_n)=\varphi(b_1\cdots b_m)$
and thus $\varphi(\bar{a}_n\cdots \bar{a}_1)=\varphi(a_1\cdots a_n)^{-1}=\varphi(b_1\cdots b_m)^{-1}=\varphi(\bar{b}_m\cdots\bar{b}_1)$
and therefore $\bar{a}_n\cdots\bar{a}_1=\bar{b}_m\cdots\bar{b}_1$. Note that
$w\bar{w}=1$ for every word $w$ over positive generators. With this definition,
the set $\bar{S}=\{\bar{s}\mid s\in S\}$ is also rational. We claim that for a
word $w\in \M\Gamma$ over positive generators, $1\in wR\bar{S}$ if and only if
$1\in \varphi(w)\mathring{R}\mathring{S}^{-1}$.

If $1\in \varphi(w)\mathring{R}\mathring{S}^{-1}$, there are
$\mathring{r}\in\mathring{R}$, $\mathring{s}\in\mathring{S}$ with
$1=\varphi(w)\mathring{r}\mathring{s}^{-1}$ and hence
$\mathring{s}=\varphi(w)\mathring{r}$. Thus, we can find $s\in S$ and $r\in R$
with $\varphi(s)=\varphi(w)\varphi(r)$. The injectivity of $\varphi$ on words
over positive generators yields $s=wr$ and thus $1=wr\bar{s}$. Hence $1\in wR\bar{S}$.

If $1\in wR\bar{S}$, we have $1=wr\bar{s}$ for some $r\in R$ and $s\in S$. This
implies $1=\varphi(w)\varphi(r)\varphi(s)^{-1}$ and since $\varphi(r)\in
\mathring{R}$ and $\varphi(s)^{-1}\in \mathring{S}^{-1}$, we have
$1\in\varphi(w)\mathring{R}\mathring{S}^{-1}$.

Thus, given $n\in\N$, it is undecidable whether $1\in a_u^n R\bar{S}$. Now, we
construct a valence automaton over $\M\Gamma$ that reads a word $a^n$ while
multiplying $a_u$ in the storage for each input symbol and then
nondeterministically multiplies an element from $R$ and then an element from
$\bar{S}$. It accepts if and only if $1\in a_u^n R\bar{S}$.  Therefore, the
automaton accepts an undecidable unary language.
\end{proof}

We are now in a position to show the first main result of this section.  Note
that the first condition of the following theorem is similar to conditions (2)
and (3) in Theorem \ref{thm:cf} (and \ref{thm:gpgroup}): instead of
FRI-monoids (finite groups) we have looped vertices and instead of
non-FRI-monoids (infinite groups), we have unlooped vertices.
\newcommand{\scale}{0.7}
\begin{theorem}
All languages in $\VAl{\M\Gamma}$ are semilinear if and only if
\begin{thmlist}
\item\label{semicond:subgraphs} $\Gamma$ contains neither $\bpair{\scale}$ nor
$\bztriangle{\scale}$ as an induced subgraph and
\item\label{semicond:usubgraphs} $\Gamma$'s underlying loop-free graph contains
neither \Cfour{} nor \Pfour{} as an induced subgraph.
\end{thmlist}
\end{theorem}
\begin{proof}
Let $\Gamma=(V,E)$. Suppose conditions \ref{semicond:subgraphs} and
\ref{semicond:usubgraphs} hold. We proceed by induction on $|V|$.
\ref{semicond:usubgraphs} implies that $\Gamma$'s underlying loop-free graph is
a transitive forest. If $\Gamma$ is not connected, then $\M\Gamma$ is a free
product of graph products $\M\Gamma_1$ and $\M\Gamma_2$, for which
$\VAl{\M\Gamma_i}$ contains only semilinear languages by induction. Hence, by
Theorems \ref{thm:algsemilinear} and \ref{thm:amalgalgebraic}, every language
in $\VAl{\M\Gamma}$ is semilinear. If $\Gamma$ is connected, there is a vertex
$v\in V$ that is adjacent to every vertex other than itself. We distinguish two
cases.
\begin{itemize}
\item If $v$ is a looped vertex, then
$\VAl{\M\Gamma}=\VAl{\Z\times\M(\Gamma\setminus\{v\})}$, which contains only
semilinear languages by induction and Lemma \ref{lemma:prodz}.
\item If $v$ is an unlooped vertex, then by \ref{semicond:subgraphs},
$V\setminus\{v\}$ induces a clique of looped vertices.  Thus,
$\M\Gamma\cong\B\times\Z^{|V|-1}$, meaning $\VAl{\M\Gamma}$ contains only
semilinear languages by Lemma \ref{lemma:prodz}.
\end{itemize}

We shall now prove the other direction. If $\Gamma$ contains $\bpair{\scale}$
as an induced subgraph, then $\VAl{\B\times\B}$ is included in $\VAl{\M\Gamma}$
and the former contains a non-semilinear language by Lemma
\ref{lemma:bbnonsemilinear}.  If $\Gamma$ contains $\bztriangle{\scale}$, then
$\M\Gamma$ contains a copy of $\B\times (\Z*\Z)$ as a submonoid. By Theorem \ref{thm:freecf},
we have $\VAl{\B}\subseteq\VAl{\Z*\Z}$ and hence Lemma
\ref{lemma:prodinclusion} implies $\VAl{\B\times\B}\subseteq\VAl{\B\times
(\Z*\Z)}$. 

Suppose $\Gamma$'s underlying loop-free graph contains \Cfour{} as an induced
subgraph.  Since we have already shown that the presence of $\bpair{\scale}$ or
$\bztriangle{\scale}$ as an induced subgraph guarantees a non-semilinear
language in $\VAl{\M\Gamma}$, we may assume that all four participating
vertices are looped. Hence, $\M\Gamma$ contains a copy of
$(\Z*\Z)\times(\Z*\Z)$. By Theorem \ref{thm:freecf} and Lemma
\ref{lemma:prodinclusion}, this means
$\VAl{\B\times\B}\subseteq\VAl{\M\Gamma}$. Thus, $\VAl{\M\Gamma}$ contains a
non-semilinear language.  Finally, if $\Gamma$'s underlying loop-free graph
contains \Pfour{} as an induced subgraph, Lemma \ref{lemma:pfour} provides the
existence of an undecidable unary language in $\VAl{\M\Gamma}$.  Since such a
language cannot be semilinear, the lemma is proven.
\end{proof}

\subsection{Torsion groups}

A \emph{torsion group} is a group $G$ in which for each $g\in G$, there is a
$k\in\N\setminus\{0\}$ with $g^k=1$. In this subsection, we show that for
torsion groups $G$, all languages in $\VAl{G}$ are semilinear. The key
ingredient in our proof is showing that a certain set of multisets is upward
closed with respect to a well-quasi-ordering.  A \emph{well-quasi-ordering on
$A$} is a reflexive transitive relation $\le$ on $A$ such that for every
infinite sequence $(a_n)_{n\in\N}$, $a_n\in A$, there are indices $i<j$ with
$a_i\le a_j$. We call a subset $B\subseteq A$ \emph{upward closed} if $a\in B$
and $a\le b$ imply $b\in B$.  A basic observation about well-quasi-ordered sets
states that for each upward closed set $B\subseteq A$, the set of its minimal
elements is finite and $B$ is the set of those $a\in A$ with $m\le a$ for some
minimal $m\in B$ (see \cite{Kruskal1972}).

Given multisets $\alpha,\beta\in X^\oplus$ and $k\in\N$, we write
$\alpha\equiv_k\beta$ if $\alpha(x)\equiv \beta(x) \pmod{k}$ for each $x\in X$.
Furthermore, we write $\alpha\le_k\beta$ if $\alpha\le\beta$ and
$\alpha\equiv_k\beta$. Clearly, $\le_k$ is a well-quasi-ordering on $X^\oplus$:
Since $\equiv_k$ has finite index in $X^\oplus$, we find in any infinite
sequence $\alpha_1,\alpha_2,\ldots\in X^\oplus$ an infinite subsequence
$\alpha'_1,\alpha'_2,\ldots\in X^\oplus$ of $\equiv_k$-equivalent multisets.
Furthermore, $\le$ is well-known to be a well-quasi-ordering~\cite{Dickson1913} and yields indices
$i<j$ with $\alpha'_i\le\alpha'_j$ and hence $\alpha'_i\le_k\alpha'_j$.

If $S\subseteq X^\oplus$ is upward closed with respect to $\le_k$, we also say
$S$ is \emph{$k$-upward-closed}.  The observation above means in particular
that every $k$-upward-closed set is semilinear.

\begin{theorem}\label{thm:torsion}
For every torsion group $G$, the languages in $\VAl{G}$ are semilinear.
\end{theorem}
\begin{proof}
Let $G$ be a torsion group and $K$ be accepted by the valence automaton
$A=(Q,X,G,E,q_0,F)$. We regard the finite set $E$ as an alphabet and define the
automaton $\hat{A}=(Q,E,G,\hat{E},q_0,F)$ such that $\hat{E}=\{(p,(p,w,g,q),g,q)
\mid (p,w,g,q)\in E\}$.  Let $\hat{K}=L(\hat{A})$. Clearly, in order to prove
Theorem \ref{thm:torsion}, it suffices to show that $\hat{K}$ is semilinear.

\newcommand{\states}[1]{\sigma(#1)}
For a word $w\in E^*$, $w=(p_1, x_1, g_1, q_1)\cdots(p_n, x_n, g_n, q_n)$, we
write $\states{w}$ for the set $\{p_i, q_i \mid 1\le i\le n\}$.  $w$ is called
a \emph{$p,q$-computation} if $p_1=p$, $q_n=q$, and $q_i=p_{i+1}$ for $1\le
i<n$.  A $q,q$-computation is also called a \emph{$q$-loop}. Moreover, a
$q$-loop $w$ is called \emph{simple} if $q_i\ne q_j$ for $i\ne j$.

For each subset $S\subseteq Q$, let $F_S$ be the set of all words $w\in E^*$
with $\states{w}=S$ and for which there is a $q\in F$ such that $w$ is a
$q_0,q$-computation and $|w|\le |Q|\cdot (2^{|Q|}+1)$. Furthermore, let
$L_S\subseteq E^*$ consist of all $w\in E^*$ such that $w$ is a simple $q$-loop
for some $q\in S$ and $\states{w}\subseteq S$. Note that $L_S$ is finite, which
allows us to define the alphabet $Y_S$ so as to be in bijection with $L_S$.
Let $\varphi:Y_S\to L_S$ be this bijection and let $\tilde{\varphi}:
Y_S^\oplus\to E^\oplus$ be the morphism satisfying
$\tilde{\varphi}(y)=\Parikh{\varphi(y)}$ for $y\in Y_S$.

For $p,q$-computations $v,w\in E^*$, we write $v\vdash w$ if
$\states{v}=\states{w}$ and $w=rst$ such that $r$ is a $p,q'$-computation, $s$
is a simple $q'$-loop, $t$ is a $q',q$-computation, and $v=rt$. Moreover, let
$\preceq$ be the reflexive transitive closure of $\vdash$.  In other words,
$v\preceq w$ means that $w$ can be obtained from $v$ by inserting simple
$q$-loops for states $q\in Q$ without increasing the set of visited states. For
each $v\in F_S$, we define
\[ U_v = \{ \mu\in Y_S^\oplus \mid \exists w\in \hat{K}: v\preceq w,~\Parikh{w}=\Parikh{v}+\tilde{\varphi}(\mu) \} \]
(note that there is only one $S\subseteq Q$ with $v\in F_S$).  We claim that
\begin{equation} \Parikh{\hat{K}} = \bigcup_{S\subseteq Q} \bigcup_{v\in F_S} \Parikh{v}+\tilde{\varphi}(U_v). \tag{$\ast$}\label{eq:torsion:decomp}\end{equation}
The inclusion ``$\supseteq$'' holds by definition. For the other direction, we show 
by induction on $n$ that for any $q_f\in F$ and any $q_0,q_f$-computation $w\in E^*$, $|w|=n$,
there is a $v\in F_S$ for $S=\states{w}$ and a $\mu\in Y_S^\oplus$ with $v\preceq w$ and
$\Parikh{w}=\Parikh{v}+\tilde{\varphi}(\mu)$. If $|w|\le |Q|\cdot (2^{|Q|}+1)$,
this is satisfied by $v=w$ and $\mu=0$. Therefore, assume $|w|>|Q|\cdot
(2^{|Q|}+1)$ and write $w=(p_1, x_1, g_1, q_1)\cdots(p_n,x_n,g_n,q_n)$. Since
$n=|w|>|Q|\cdot (2^{|Q|}+1)$, there is a $q\in Q$ that appears more than
$2^{|Q|}+1$ times in the sequence $q_1,\ldots,q_n$. Hence, we can write
\[ w = w_0(p'_1,x'_1,g'_1,q)w_1\cdots (p'_m,x'_m,g'_m,q)w_m \]
with $m>2^{|Q|}+1$. Observe that for each $1\le i<m$, the word
$w_i(p'_{i+1},x'_{i+1},g'_{i+1},q)$ is a $q$-loop.  Since $m-1>2^{|Q|}$, there
are indices $1\le i<j<m$ with 
\[ \states{w_i(p'_{i+1},x'_{i+1},g'_{i+1},q)}=\states{w_j(p'_{j+1},x'_{j+1},g'_{j+1},q)}. \]
Furthermore, we can find a simple $q$-loop $\ell$ as a subword of
$w_i(p'_{i+1},x'_{i+1},g'_{i+1},q)$.  This means for the word $w'\in E^*$,
which is obtained from $w$ by removing $\ell$, we have $\states{w'}=\states{w}$
and thus $w'\vdash w$. Moreover, with $S=\sigma(w)$ and $\varphi(y)=\ell$, $y\in Y_S$, we have
$\Parikh{w}=\Parikh{w'}+\tilde{\varphi}(y)$.  Finally, since $|w'|<|w|$, the
induction hypothesis guarantees a $v\in F_S$ and a $\mu\in Y_S^\oplus$ with
$v\preceq w'$ and $\Parikh{w'}=\Parikh{v}+\tilde{\varphi}(\mu)$. Then we have
$v\preceq w$ and $\Parikh{w}=\Parikh{v}+\tilde{\varphi}(\mu+y)$ and the
induction is complete.  In order to prove ``$\subseteq$'' of
\eqref{eq:torsion:decomp}, suppose $w\in \hat{K}$. Since $w$ is a
$q_0,q_f$-computation for some $q_f\in F$, we can find the above $v\in F_S$,
$S=\states{w}$, and $\mu\in Y_S^\oplus$ with $v\preceq w$ and
$\Parikh{w}=\Parikh{v}+\tilde{\varphi}(\mu)$.  This means $\mu\in U_v$ and
hence $\Parikh{w}$ is contained in the right hand side of
\eqref{eq:torsion:decomp}.  This proves \eqref{eq:torsion:decomp}.

By \eqref{eq:torsion:decomp} and since $F_S$ is finite for each $S\subseteq Q$,
it suffices to show that $U_v$ is semilinear for each $v\in F_S$ and
$S\subseteq Q$.  Let $\gamma:E^*\to G$ be the morphism with
$\gamma((p,x,g,q))=g$ for $(p,x,g,q)\in E$.  Since $G$ is a torsion group, the
finiteness of $L_S$ permits us to choose a $k\in\N$ such that
$\gamma(\ell)^k=1$ for any $\ell\in L_S$. We claim that $U_v$ is
$k$-upward-closed. It suffices to show that for $\mu\in U_v$, we also have
$\mu+k\cdot y\in U_v$ for any $y\in Y_S$.  Hence, let $\mu\in U_v$ with
$w\in\hat{K}$ such that $v\preceq w$ and
$\Parikh{w}=\Parikh{v}+\tilde{\varphi}(\mu)$ and let $\mu'=\mu+k\cdot y$.  Let
$\ell=\varphi(y)\in L_S$ be a simple $q$-loop. Then $q\in S$ and since
$\sigma(w)=\sigma(v)=S$, we can write $w=r(q_1, x_1, g_1, q)s$, $r,s\in E^*$.
The fact that $w\in\hat{K}$ means in particular $\gamma(w)=1$. Therefore, the
word $w'=r(q_1, x_1, g_1, q)\ell^ks$ is a $q_0,q_f$-computation for some
$q_f\in F$ and satisfies $\gamma(w')=1$ since $\gamma(\ell)^k=1$.  This means
$w'\in \hat{K}$ and
$\Parikh{w'}=\Parikh{w}+k\cdot\Parikh{\ell}=\Parikh{v}+\tilde{\varphi}(\mu+k\cdot
y)$. We also have $\sigma(\ell)\subseteq S$ and hence $v\preceq w\preceq w'$.
Therefore, $\mu'=\mu+k\cdot y\in U_v$.  This proves $U_v$ to be
$k$-upward-closed and thus semilinear.
\end{proof}

Render \cite{Render2010} has shown that for every monoid $M$, the class
$\VAl{M}$ either (1) coincides with the regular languages, (2) contains the
blind one-counter languages, (3) contains the partially blind one-counter
languages, or (4) consists of those accepted by valence automata over an infinite
torsion group (which is not locally finite). Hence, we obtain the following.
\begin{corollary}
For each monoid $M$, at least one of the following holds:
\begin{thmlist}
\item $\VAl{M}$ contains only semilinear languages.
\item $\VAl{M}$ contains the languages of blind one-counter automata.
\item $\VAl{M}$ contains the languages of partially blind one-counter automata.
\end{thmlist}
\end{corollary}

Since there are torsion groups with an undecidable word problem \cite{Adian2010}, we have:
\begin{scorollary}
There is a group $G$ with an undecidable word problem such that all languages
in $\VAl{G}$ are semilinear.
\end{scorollary}

As another application, we can show that the one-sided Dyck language is not
accepted by any valence automaton over $G\times\Z^n$, where $G$ is a torsion
group and $n\in\N$.
\begin{scorollary}\label{cor:dyck}
For torsion groups $G$ and $n\in\N$, we have $D_1\notin\VAl{G\times\Z^n}$.
\end{scorollary}
\begin{proof}
First, observe that $\VAl{\B\times\B}$ is not contained in $\VAl{G\times\Z^n}$,
since the former contains a non-semilinear language by Lemma
\ref{lemma:bbnonsemilinear} and the latter contains only semilinear ones by
Theorem \ref{thm:torsion} and Lemma \ref{lemma:prodz}.

If $D_1$ were contained in $\VAl{G\times\Z^n}$, then
$\VAl{\B}\subseteq\VAl{G\times\Z^n}$, since $D_1$ is an identity language of
$\B$. This means that $\VAl{\B\times\B}$ is contained in the class of languages
accepted by valence automata over $(G\times\Z^n)\times(G\times\Z^n)$. The
latter group, however, is isomorphic to $G^2\times\Z^{2n}$, contradicting our
observation above.
\end{proof}

\paragraph*{Acknowledgements} We are indebted to one of the anonymous referees
for MFCS 2013, who pointed out a misuse of terminology in a previous version of
Theorem \ref{thm:amalgalgebraic}.

\bibliographystyle{plain}
\bibliography{bibliography}

\end{document}